\documentclass[twocolumn,amsmath,prb,superscriptaddress]{revtex4}
\pdfoutput=1

\usepackage{graphicx}

\begin{document}

\title{Investigation of Anti-Relaxation Coatings for Alkali-Metal Vapor Cells Using Surface Science Techniques}

\author{S.~J.~Seltzer\footnote{These authors have made equal contributions to this work.}}
\email{seltzer@berkeley.edu}
\affiliation{Materials Sciences Division, Lawrence Berkeley National Laboratory, Berkeley, California 94720}
\affiliation{Department of Chemistry, University of California, Berkeley, California 94720}

\author{D.~J.~Michalak\footnotemark[1]}
\altaffiliation{Present address: Components Research, Intel Corporation, Hillsboro, Oregon 97124}
\affiliation{Materials Sciences Division, Lawrence Berkeley National Laboratory, Berkeley, California 94720}
\affiliation{Department of Chemistry, University of California, Berkeley, California 94720}

\author{M.~H.~Donaldson}
\affiliation{Materials Sciences Division, Lawrence Berkeley National Laboratory, Berkeley, California 94720}
\affiliation{Department of Chemistry, University of California, Berkeley, California 94720}

\author{M.~V.~Balabas}
\affiliation{S. I. Vavilov State Optical Institute, St. Petersburg, 199034 Russia}

\author{S.~K.~Barber}
\affiliation{Advanced Light Source, Lawrence Berkeley National Laboratory, Berkeley, California 94720}

\author{S.~L.~Bernasek}
\affiliation{Department of Chemistry, Princeton University, Princeton, New Jersey 08544}

\author{M.-A.~Bouchiat}
\affiliation{Laboratoire Kastler Brossel, D\'{e}partement de Physique de l'Ecole Normale Sup\'{e}rieure, F-75231 Paris, France}

\author{A.~Hexemer}
\affiliation{Advanced Light Source, Lawrence Berkeley National Laboratory, Berkeley, California 94720}

\author{A.~M.~Hibberd}
\affiliation{Department of Chemistry, Princeton University, Princeton, New Jersey 08544}

\author{D.~F.~Jackson~Kimball}
\affiliation{Department of Physics, California State University-East Bay, Hayward, California 94542}

\author{C.~Jaye}
\affiliation{Ceramics Division, Materials Science and Engineering Laboratory, National Institute of Standards and Technology, Gaithersburg, Maryland 20899}

\author{T.~Karaulanov}
\affiliation{Department of Physics, University of California, Berkeley, California 94720}

\author{F.~A.~Narducci}
\affiliation{EO Sensors Division, US Naval Air Systems Command, Patuxent River, Maryland 20670}

\author{S.~A.~Rangwala}
\affiliation{Raman Research Institute, Sadashivanagar, Bangalore 560080 India}

\author{H.~G.~Robinson}
\affiliation{Time and Frequency Division, National Institute of Standards and Technology, Boulder, Colorado 80305}

\author{A.~K.~Shmakov}
\affiliation{Department of Physics, University of California, Berkeley, California 94720}

\author{D.~L.~Voronov}
\affiliation{Advanced Light Source, Lawrence Berkeley National Laboratory, Berkeley, California 94720}

\author{V.~V.~Yashchuk}
\affiliation{Advanced Light Source, Lawrence Berkeley National Laboratory, Berkeley, California 94720}

\author{A.~Pines}
\affiliation{Materials Sciences Division, Lawrence Berkeley National Laboratory, Berkeley, California 94720}
\affiliation{Department of Chemistry, University of California, Berkeley, California 94720}

\author{D.~Budker}
\email{budker@berkeley.edu}
\affiliation{Nuclear Science Division, Lawrence Berkeley National Laboratory, Berkeley, California 94720}
\affiliation{Department of Physics, University of California, Berkeley, California 94720}

\begin{abstract}
Many technologies based on cells containing alkali-metal atomic vapor benefit from the use of anti-relaxation surface coatings in order to preserve atomic spin polarization. In particular, paraffin has been used for this purpose for several decades and has been demonstrated to allow an atom to experience up to 10,000 collisions with the walls of its container without depolarizing, but the details of its operation remain poorly understood. We apply modern surface and bulk techniques to the study of paraffin coatings, in order to characterize the properties that enable the effective preservation of alkali spin polarization. These methods include Fourier transform infrared spectroscopy, differential scanning calorimetry, atomic force microscopy, near-edge X-ray absorption fine structure spectroscopy, and X-ray photoelectron spectroscopy. We also compare the light-induced atomic desorption yields of several different paraffin materials. Experimental results include the determination that crystallinity of the coating material is unnecessary, and the detection of C=C double bonds present within a particular class of effective paraffin coatings. Further study should lead to the development of more robust paraffin anti-relaxation coatings, as well as the design and synthesis of new classes of coating materials.
\end{abstract}

\maketitle

\section{Introduction}

Paraffin films and other surface coatings play an integral part in several emerging technologies that employ vapors of alkali-metal atoms, including atomic magnetometers, clocks, and quantum and nonlinear optical devices. Paraffin was first shown to preserve the spin polarization of alkali atoms in 1958 by Robinson, Ensberg, and Dehmelt \cite{RobinsonBulletin} and was first studied extensively by Bouchiat and Brossel.\cite{BouchiatThesis, BouchiatBrossel}  It has been investigated by several others in the decades since,\cite{RobinsonAPL, LibermanKnize, GrafParaffin, CastagnaParaffin} but the details of its operation as an anti-relaxation coating remain poorly understood. Paraffin coatings enable narrow Zeeman resonance linewidths, and they have recently been the subject of renewed interest due to advances in the technology of alkali-metal magnetometers \cite{BudkerRomalis, WasilewskiEntanglement, ShahQND} that have led to the development of detectors with sensitivity comparable to or better than superconducting quantum interference devices (SQUIDs). Modern magnetometers have enabled significant advances in low-magnetic-field nuclear magnetic resonance (NMR),\cite{XuRev, SavukovRFNMR, LedbetterPNAS} magnetic resonance imaging (MRI),\cite{XuRev, XuPNAS, XuPRA, XuJMR, SavukovJMR} and medical imaging,\cite{WeisHeart, XiaMEG, BelfiCPT} as well as paleomagnetism,\cite{Dang} explosives detection,\cite{LeeNQR} and ultra-sensitive tests of fundamental physics.\cite{BerglundLimits, VasilakisLimits} Paraffin-coated cells also feature narrow hyperfine resonance linewidths and have been explored in the context of secondary frequency standards;\cite{BudkerFreq, GuzmanFreq} they have also been employed in experiments involving spin squeezing,\cite{KuzmichSqueezing} quantum memory,\cite{JulsgaardMemory} and ``slow light.''\cite{KleinSlow} While cells with diameters from a few to tens of centimeters are typically employed, miniature millimeter-sized cells with paraffin coating have also been explored.\cite{BalabasJOpt} In addition, similar coatings with silicon head groups have been used in magneto-optical traps,\cite{StephensTraps, LuTraps, GuckertTraps, GomezSpectroscopy} hollow photonic fibers,\cite{GhoshPhotonic, LightPhotonic} and noble-gas-atom optical pumping cells.\cite{ZengXe, DriehuysXe}

Alkali atoms in the vapor phase depolarize upon contact with the bare surface of a glass container, limiting the coherence lifetime of the spin ensemble. In order to prevent such depolarization, vapor cells typically include either a buffer gas or an anti-relaxation surface coating. The inclusion of up to several atmospheres of a chemically inert buffer gas slows diffusion of alkali atoms to the cell walls, but there are several advantages to the use of a surface coating, including lower laser power requirements, larger optical signals, reduced influence of magnetic-field gradients, and a smaller collision rate with other atoms and molecules in the cell. Anti-relaxation coatings can allow an alkali atom to experience thousands of collisions with the walls of the cell without depolarizing, and paraffin in particular has been demonstrated to allow up to 10,000 bounces.\cite{BouchiatBrossel} As the size of the vapor cell decreases, the surface-area-to-volume ratio increases, requiring improvement in the quality of the surface coating to compensate for the resulting increase in the rate of collisions with the surface, in order to maintain the same spin-coherence lifetime. In miniature alkali vapor cells with volume 1-10~mm$^{3}$ or less, the use of surface coatings enables more sensitive magnetometers and clocks than the use of buffer gas, assuming that a coating with appropriate spin-preservation properties can be employed.\cite{KitchingMiniature}

A paraffin coating significantly reduces the probability of spin destruction during a collision with the surface because it contains no free electron spins and it features a lower adsorption energy than the bare glass, thus reducing the residence time of an adsorbed alkali atom on the surface of the paraffin relative to the glass surface.  The residence time of an atom at a surface site can be expressed as $\tau_{r}=\tau_{0}\exp\left( E_{a}/kT \right)$, where $E_{a}$ is the adsorption energy, $k$ is the Boltzmann constant, and $T$ is the temperature.\cite{BouchiatBrossel} $\tau_{0}$ is the time constant for a perfectly elastic collision and therefore gives the high-temperature limit where the thermal energy $kT\gg E_{a}$.  Minimal $\tau_{r}$ is desirable because an adsorbed alkali atom dephases from the ensemble of atoms in the bulk of the cell, as a result of experiencing both a different magnetic field than in the cell interior and a fluctuating magnetic field generated by the hydrogen nuclei of the paraffin material.\cite{BouchiatBrossel} The adsorption energy for alkali atoms on a paraffin surface is small, roughly 0.06-0.1~eV,\cite{BouchiatBrossel, RahmanJQE, BudkerFreq} and assuming $\tau_{0}$$\sim$10$^{-12}$~s (the period of a typical molecular vibration) gives a residence time of approximately 0.1~ns at room temperature.

The performance of paraffin [C$_{n}$H$_{2n+2}$] coatings quickly degrades at temperatures above 60-80$^{\circ}$C,\cite{BouchiatBrossel, RahmanJQE} but operation at higher temperature is beneficial for some devices because it increases the saturated vapor pressure of the alkali atoms, and thus the atomic density. For most types of paraffin, such as tetracontane ($n$=40), the critical temperature corresponds to the melting point,\cite{RahmanJQE} but for longer-chain polyethylene coatings,\cite{BouchiatBrossel} which have a much higher fusion temperature of 130$^{\circ}$C, the mechanism for the decreased performance above 60$^{\circ}$C is not fully understood. Recent magnetometers have achieved ultra-high sensitivity better than 1~fT/$\sqrt{\textnormal{Hz}}$ to near-dc \cite{KominisNature} and radio-frequency \cite{SavukovRF} magnetic fields by operating at very high vapor density, but the high operating temperatures of these magnetometers ($T$$>$100$^{\circ}$C for cesium vapor and $T$$>$150$^{\circ}$C for potassium vapor) prevent the use of paraffin coatings. In addition, paraffin does not survive the elevated temperatures required by the anodic bonding process used in the production of microfabricated vapor cells. Surface coatings with superior temperature stability are therefore required for use with high-density or microfabricated alkali vapor cells. High-temperature coatings also allow experimentation with potassium and sodium vapor, which have lower vapor pressures compared to rubidium and cesium at a given temperature.

Recent efforts at developing alternatives to paraffin have mainly focused on certain silane coatings that resemble paraffin, containing a long chain of hydrocarbons but also a silicon head group that chemically binds to the glass surface.\cite{StephensTraps, YiMonolayers, ZhaoPyrex, SeltzerReusable, Rampulla} Such materials do not melt and remain attached to the glass surface at relatively high temperatures, enabling them to function as anti-relaxation coatings at much higher temperatures than paraffin. In particular, a multilayer coating of octadecyltrichlorosilane [OTS, CH$_{3}$(CH$_{2}$)$_{17}$SiCl$_{3}$] has been observed to allow from hundreds up to 2100 bounces with the cell walls \cite{SeltzerBeats} and can operate in the presence of potassium and rubidium vapor up to about 170$^{\circ}$C.\cite{SeltzerRomalis} However, the quality of such coatings with respect to preserving alkali polarization is highly variable, even between cells coated in the same batch, and remains significantly worse than that achievable with paraffin. In addition, it was shown recently that an alkene coating, which resembles paraffin except with an unsaturated C=C double bond, can allow up to two orders of magnitude more bounces than paraffin, but only to temperatures of 33$^{\circ}$C,\cite{BalabasHighQuality, BalabasMagic} and properties such as stability of the coating remain to be studied.

Paraffin thus remains the most widely used anti-relaxation coating for alkali spins. In order to facilitate the design and development of new coating materials for alkali-metal cells and hollow fibers, it is therefore necessary to develop a more detailed understanding of the interactions between alkali atoms and the paraffin surface, many aspects of which are not yet fully understood. Indeed, the production of high-quality paraffin cells remains more of an art than a science, with little understanding of why only certain coating procedures work or the reasons that variations in those procedures affect the anti-relaxation quality of the coating. As an example, paramagnetic impurities could couple to and depolarize the alkali spins, so some researchers observe that purification of the paraffin by distillation is necessary to produce a high-quality coating;\cite{RobinsonPersonal} however, others have been able to use paraffins as received.\cite{BouchiatBrossel}  In addition, a so-called ``ripening'' phase is required, which involves annealing of the paraffin-coated cell at 50-60$^{\circ}$C in the presence of the alkali metal for an extended period of time (typically hours to days) before the cell may be used, although the specific processes that occur during ripening remain unknown.

Much of the behavior of the paraffin coating during operation remains equally mysterious. For example, the paraffin coating introduces a small shift in the hyperfine frequency of the atomic ground state, which varies between cells and must be accounted for in the use of alkali vapor cells as frequency standards.\cite{Vanier2, BudkerFreq} In addition, the measured vapor pressure in a paraffin-coated cell is smaller than the expected saturated vapor pressure at the temperature of the cell, implying that alkali atoms can be absorbed into, and thus can diffuse within, the bulk volume of the paraffin.\cite{BouchiatThesis, LibermanKnize, BalabasPrzh} It has been observed \cite{AleksandrovLIAD,GozziniSodium} that alkali vapor density increases significantly when a paraffin-coated cell is exposed to light (particularly uv and near-uv light) due to the light-induced atomic desorption (LIAD) effect,\cite{GozziniLIAD} which causes absorbed atoms to be ejected from the paraffin coating.  Recently, LIAD effects on alkali spin relaxation have been investigated,\cite{GrafParaffin} with a focus on non-thermal control of alkali vapor density.\cite{KaraulanovThermal} The exact mechanisms of LIAD in paraffin are not yet understood; however, it is clear that a combination of surface processes and light-enhanced diffusion of alkali atoms within the bulk of the coating are important. LIAD has been used with silane coatings to load photonic fibers,\cite{GhoshPhotonic} and it has been used with bare pyrex surfaces to load a chip-scale Bose-Einstein condensate (BEC),\cite{DuBEC} although in the latter case the desorption efficiency might be increased by orders of magnitude with a paraffin or silane coating. Any coating used for LIAD loading of an ultra-cold atom chip would require compatibility with ultra-high vacuum. Similarly, enhanced alkali-atom density is also observed upon the application of large electric fields (1-8~kV/cm) across paraffin-coated cells,\cite{BudkerElec, KimballElec} although again the exact mechanism remains unknown.

In this work we use modern surface science methods as the basis for an investigation of the interaction of alkali-metal atoms with various paraffin materials. Unlike most previous studies, which primarily measured properties of the alkali vapor such as density and relaxation time, we instead observe the properties of the coating itself. Fourier transform infrared spectroscopy (FTIR) and differential scanning calorimetry (DSC) were used to understand bulk properties of the coating materials, while atomic force microscopy (AFM), near edge X-ray absorption fine-structure (NEXAFS), and X-ray photoelectron spectroscopy (XPS) were used to obtain information about the coated surface and its interaction with the alkali atoms. These and similar techniques have been employed previously to study paraffins,\cite{FlahertyDSC, Casal, PlompMorphology, ZhengRaman, Tzvetkov} but not in the context of their use with alkali atoms. In addition, we compared the LIAD yields of several different paraffin materials. An array of techniques enables a more complete characterization of the atom-surface interactions than can otherwise be achieved. The work described here is intended to demonstrate the power of these methods to thoroughly characterize and help understand the behavior of paraffin and other coating materials, in order to guide the creation of new coatings to enhance the performance of rapidly developing technologies such as microfabricated magnetometers and clocks, nonlinear and quantum optical devices, and portable LIAD-loaded devices.

\section{Experimental Methods and Results}

We selected paraffin waxes that have been successfully implemented as anti-relaxation coatings in alkali vapor cells, including the $n$-alkanes eicosane [CH$_{3}$(CH$_{2}$)$_{18}$CH$_{3}$], dotriacontane [CH$_{3}$(CH$_{2}$)$_{30}$CH$_{3}$], and tetracontane [CH$_{3}$(CH$_{2}$)$_{38}$CH$_{3}$], as well as long-chain polyethylene [CH$_{3}$(CH$_{2}$)$_{n}$CH$_{3}$ where $n$ is large and varies between molecules]. We also considered a proprietary paraffin from Dr.~Mikhail Balabas that is used in the manufacture of magnetometer cells, which we refer to here as pwMB; this wax is obtained by fractionation of polyethelyene wax at 220$^{\circ}$C.\cite{AleksandrovLIAD} Finally, we considered the commercially available waxes FR-130 parowax (from the Luxco Wax company) and paraflint, which are both expected to contain a mixture of various $n$-alkanes. Unless otherwise noted, samples were coated following the typical procedures used in the manufacture of alkali vapor cells, which involve evaporating the material at high temperature in vacuum and allowing it to condense on the inner surface of the cell.

\subsection{Fourier transform infrared spectroscopy}

Fourier transform infrared spectroscopy (FTIR) analysis was performed on several types of paraffin in order to identify the general functional groups present in the waxes. FTIR is used extensively for structure determination and chemical analysis because it gives bond-specific information.\cite{InstrumAnal} Transitions due to specific bending and stretching modes of bonds absorb infrared light at specific frequencies allowing for identification of the bonds present.

Spectra were obtained on a Varian 640-IR system using a room-temperature DTGS (deuterated triglycine sulfate) detector. The aperture of the instrument was completely open, and a nominal 4~cm$^{-1}$ spectral resolution was used.  The strong optical absorption of the Si-O-Si structure makes glass substrates opaque to infrared analysis below 2000~cm$^{-1}$.  To widen the range of FTIR characterization, native oxide terminated, 500~$\mu$m thick, (111)-oriented, low-doped (phosphorus, 74-86~$\Omega\cdot$cm resistivity) silicon wafers (Silicon Quest International) were instead used as substrates for paraffin coating. The silicon phonon absorption occurs below 600~cm$^{-1}$ and therefore allows for a more thorough characterization of the various waxes.  In addition, the surface of the silicon wafer, as received, contains a very thin layer of oxidized silicon, which is assumed to be similar enough to the surface of glass cells that it does not disturb the behavior of the wax. A Si thickness of 500~$\mu$m or greater is necessary to achieve spectra with 4~cm$^{-1}$ resolution without complication due to interference from ghost zero-path difference  (etaloning effect) peaks. High-resistivity Si improves signal to noise by minimizing free-carrier absorption of the IR light.  Uncoated Si samples were first situated in the beam path with the surface plane perpendicular to the light propagation direction, and 256 background spectra of the uncoated silicon wafers were obtained in transmission mode before the waxes were rub-coated onto the warmed surface, giving a layer with thickness of hundreds of nm to several $\mu$m. To minimize baseline drift, 256 scans were collected in transmission mode under the exact same sample placement as for background spectra collection.

\begin{figure}
\centering
\includegraphics[width=8.0cm]{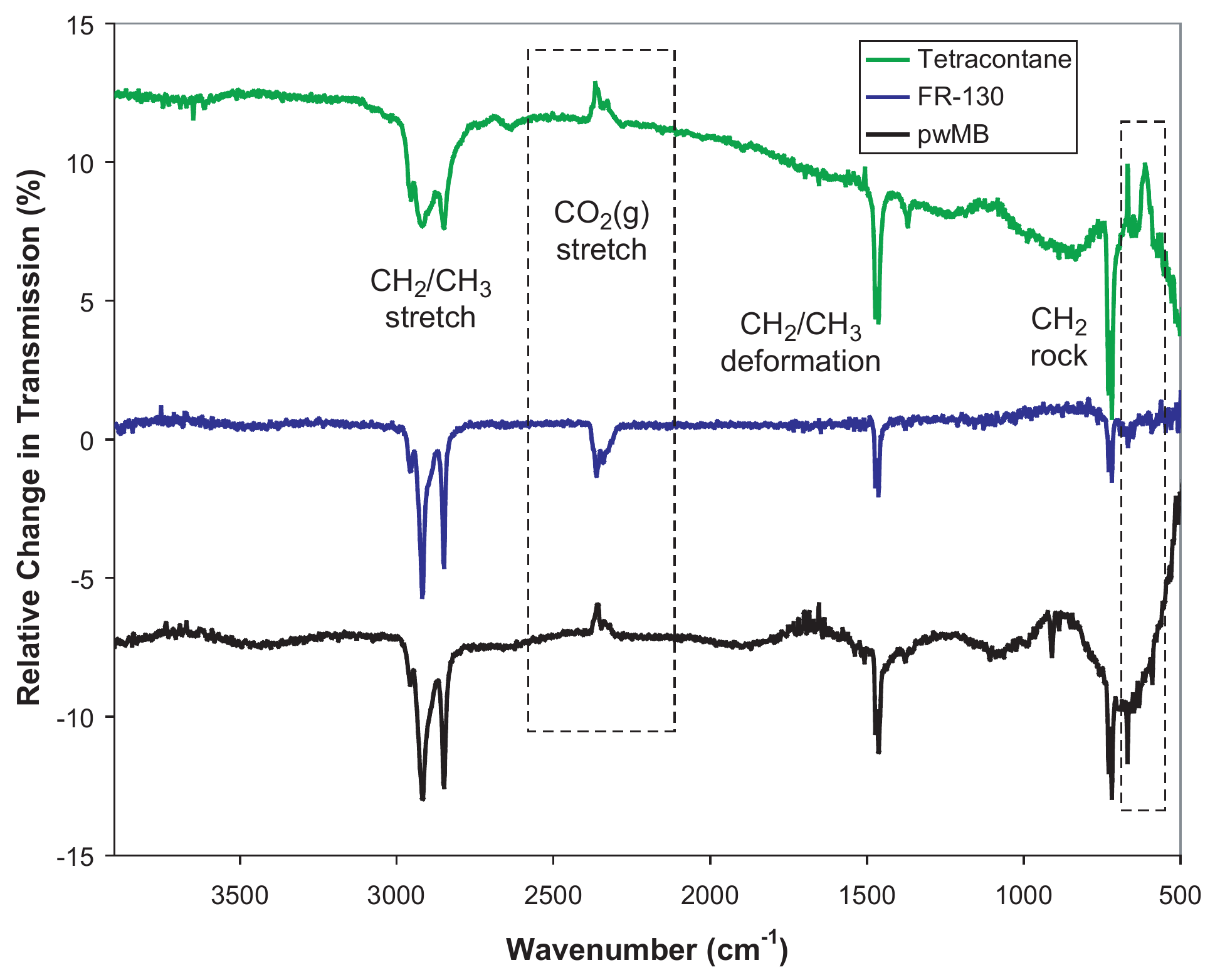}
\caption[FTIR spectra of various waxes.]{FTIR spectra of tetracontane (green, top), FR-130 (blue, middle), and pwMB (black, bottom), with the traces offset vertically for clarity. The dashed rectangles outline the peaks due to atmospheric CO$_{2}$ absorption and/or Si phonon absorption at 600~cm$^{-1}$.} \label{fig_FTIR}	
\end{figure}

The transmisison spectra for tetracontane, FR-130, and pwMB are shown in Fig.~\ref{fig_FTIR}. The spectrum for tetracontane is in agreement with literature reports: CH$_{2}$ and CH$_{3}$ stretching modes at 2850, 2920, and 2954~cm$^{-1}$, an HCH scissor at 1474~cm$^{-1}$, a CH$_{3}$ asymmetric bending mode at 1464~cm$^{-1}$, a U+W (methyl symmetric and methylene wagging) mode at 1370~cm$^{-1}$, and a methylene rocking doublet at 719 and 729~cm$^{-1}$.\cite{Casal} Similar FTIR spectra were obtained for the FR-130 and pwMB paraffins. The peaks contained within broken-line boxes are due to environmental changes that occurred between background and sample collection.  The peaks for CO$_{2}$ at 2300~cm$^{-1}$ and 676~cm$^{-1}$ may be positive or negative depending on the extent of bench purging with N$_{2}$. The broad peak at 600~cm$^{-1}$ is due to the Si bulk crystal which can be either positive or negative due to slight changes in sample temperature or angle of incidence within the bench. The periodic undulations in the pwMB spectrum are due to the thickness of the film creating an etalon effect. These results show no observable carbon-carbon double bonds within the sensitivity of the experiment; specifically, there is no detectable C=C double bond stretching in the range of 1600-1700~cm$^{-1}$, and there is no detectable =C-H stretching mode within the range of 3050-3100~cm$^{-1}$.

\subsection{Differential scanning calorimetry}

Differential scanning calorimetry (DSC) was used to assess phase transitions associated with crystallinity of the bulk paraffins. DSC measures differences in heat flow into a sample and a reference as a function of temperature as the two are subjected to controlled temperature changes. It can be used to characterize crystallization behavior and assess sample purity by analysis of the heat flow behavior near phase transitions.\cite{InstrumAnal}

\begin{figure}
\centering
\includegraphics[width=8.0cm]{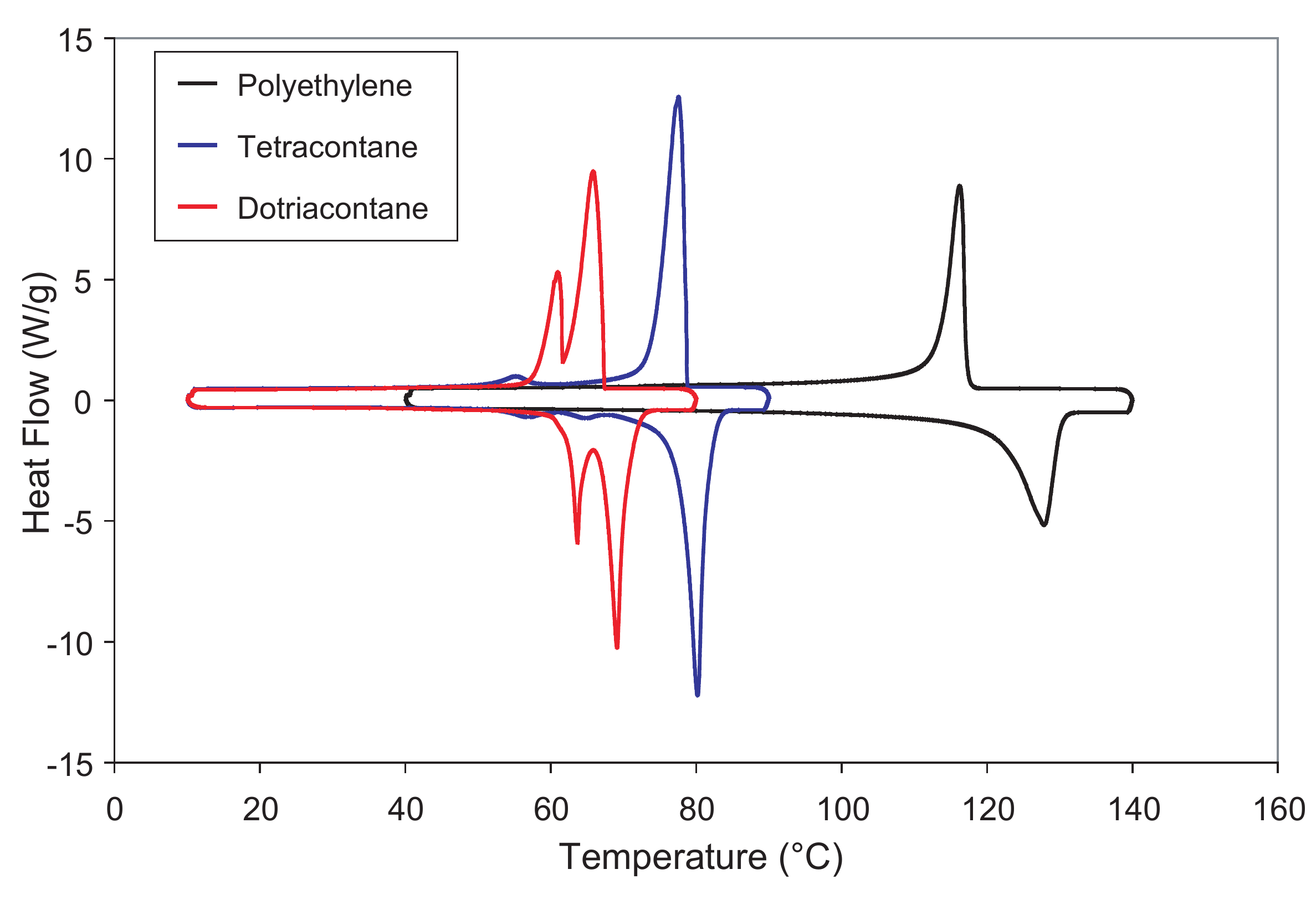}
\caption[DSC data for various monodisperse waxes.]{DSC of polyethylene (black), purified tetracontane (blue), and dotriacontane (red). The sharp peaks are indicative of crystalline phase transitions.} \label{fig_DSC1}	
\end{figure}

The DSC scans were obtained on a TA Q200-1305 Advantage instrument. Wax samples of 1-4~mg were pressed into aluminum pans and the reference was an empty aluminum pan. Measurements were made in a nitrogen atmosphere.  The typical scanning conditions were temperature scanning at a rate of 10 $^\circ$C/min over a range of 10$^\circ$C to 90-140$^\circ$C, depending on the melting point of the wax. Figure~\ref{fig_DSC1} shows the observations for several waxes expected to be crystalline at typical cell operating temperatures below 60$^{\circ}$C. Pure linear-chain alkanes (including eicosane, dotriacontane, and tetracontane) all display relatively sharp peaks indicative of the crystalline phase transitions expected for pure $n$-alkanes. Purified tetracontane displayed melting and fusion peaks at 79$^{\circ}$C, as well as a phase transition between orthorhombic and hexagonal crystal structure at 62-63$^{\circ}$C.\cite{Puchkovska} Dotriacontane similarly showed distinct melting/fusion and phase transition peaks. The long-chain polyethylene, while not completely monodisperse, showed a relatively sharp melting point extrapolated to 122$^{\circ}$C and a fusion peak at 118$^{\circ}$C, as well as a weak endothermic process during heating at 100$^{\circ}$C with a corresponding exothermic process during cooling at 80$^{\circ}$C.

\begin{figure}
\centering
\includegraphics[width=8.0cm]{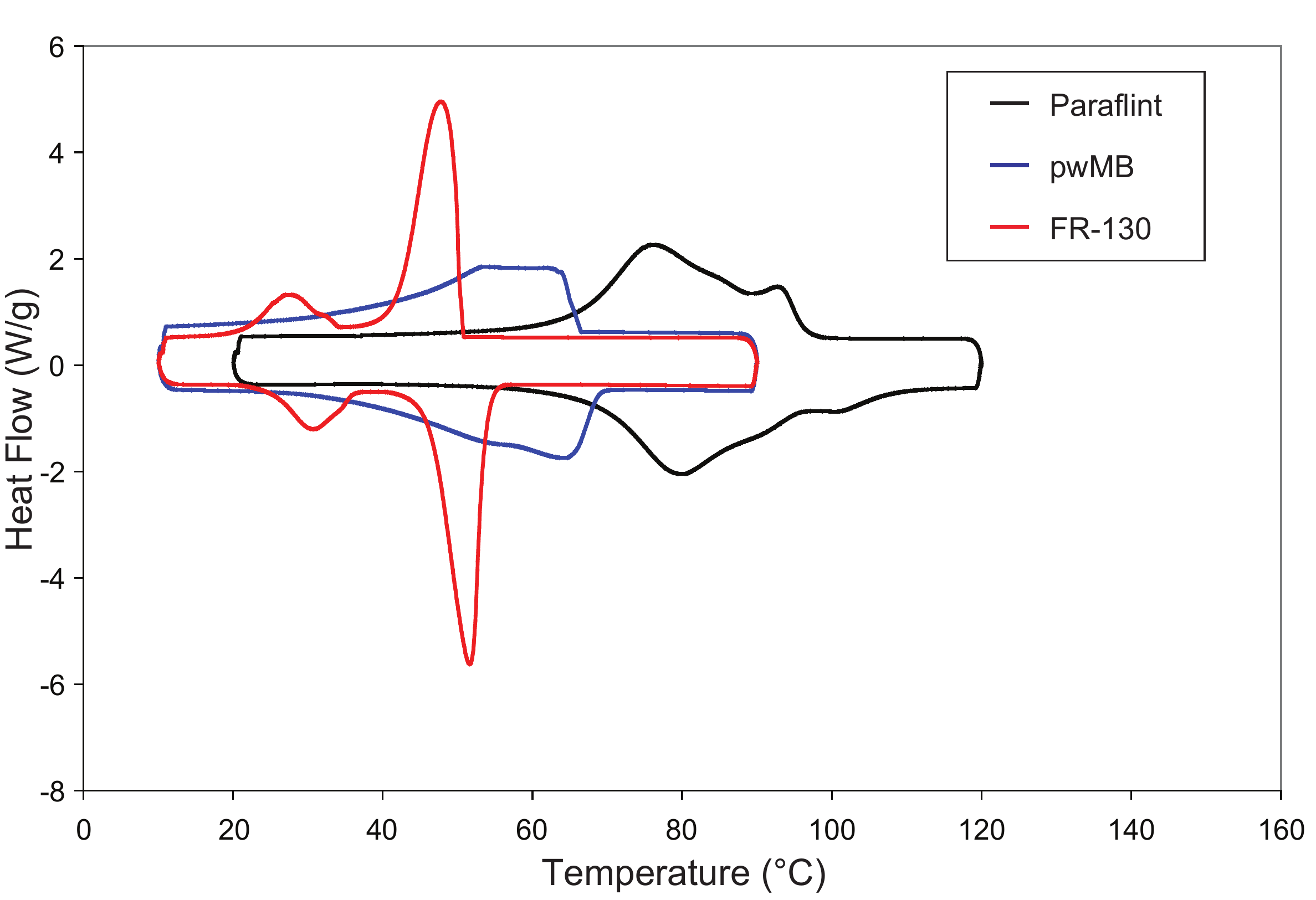}
\caption[DSC data polydisperse waxes.]{DSC of paraflint (black), pwMB (blue), and FR-130 parowax (red). The endothermic and exothermic profiles of paraflint and pwMB are very broad, indicative of a wide distribution of $n$-alkanes; these waxes are not expected to be crystalline at typical operating temperatures.} \label{fig_DSC2}	
\end{figure}

For comparison, Fig.~\ref{fig_DSC2} displays the DSC scans for several waxes that apparently do not exist in a crystalline state at typical operating temperatures. The FR-130 parowax displayed a relatively sharp melting point at 48-51$^{\circ}$C and phase transition between 25-35$^{\circ}$C, implying that this wax is fairly monodisperse.\cite{Basson} In contrast, the pwMB wax and paraflint do not display sharp melting or fusion peaks, indicating a lack of homogeneous bulk crystallinity. Instead, they feature drawn-out melting and fusion profiles that are consistent with films containing a mixture of various saturated $n$-alkanes;\cite{Basson} it is possible that these waxes contain branched alkanes as well. In particular, maximum operational temperatures of pwMB wax around 60$^{\circ}$C are well within the material's phase transition. These results indicate that crystallinity in the bulk is not necessary for an effective anti-relaxation coating, consistent with observations that some working coatings are often partially melted at standard operating temperatures.

\subsection{Atomic force microscopy}

Atomic force microscopy (AFM) \cite{BinnigAFM} of the surface of coated operational cells and coated silicon surfaces was performed to investigate the surface topography of the paraffins. AFM has been employed in the study of organic systems, especially polymeric films on solid inorganic substrates.\cite{MartiAFMorganic, MeyerAFMfilms} Measurements were taken using a DI Dimension 3100 scanning probe microscope using silicon probes with a nominal tip radius of 7~nm.

\begin{figure}
\centering
\includegraphics[width=8.0cm]{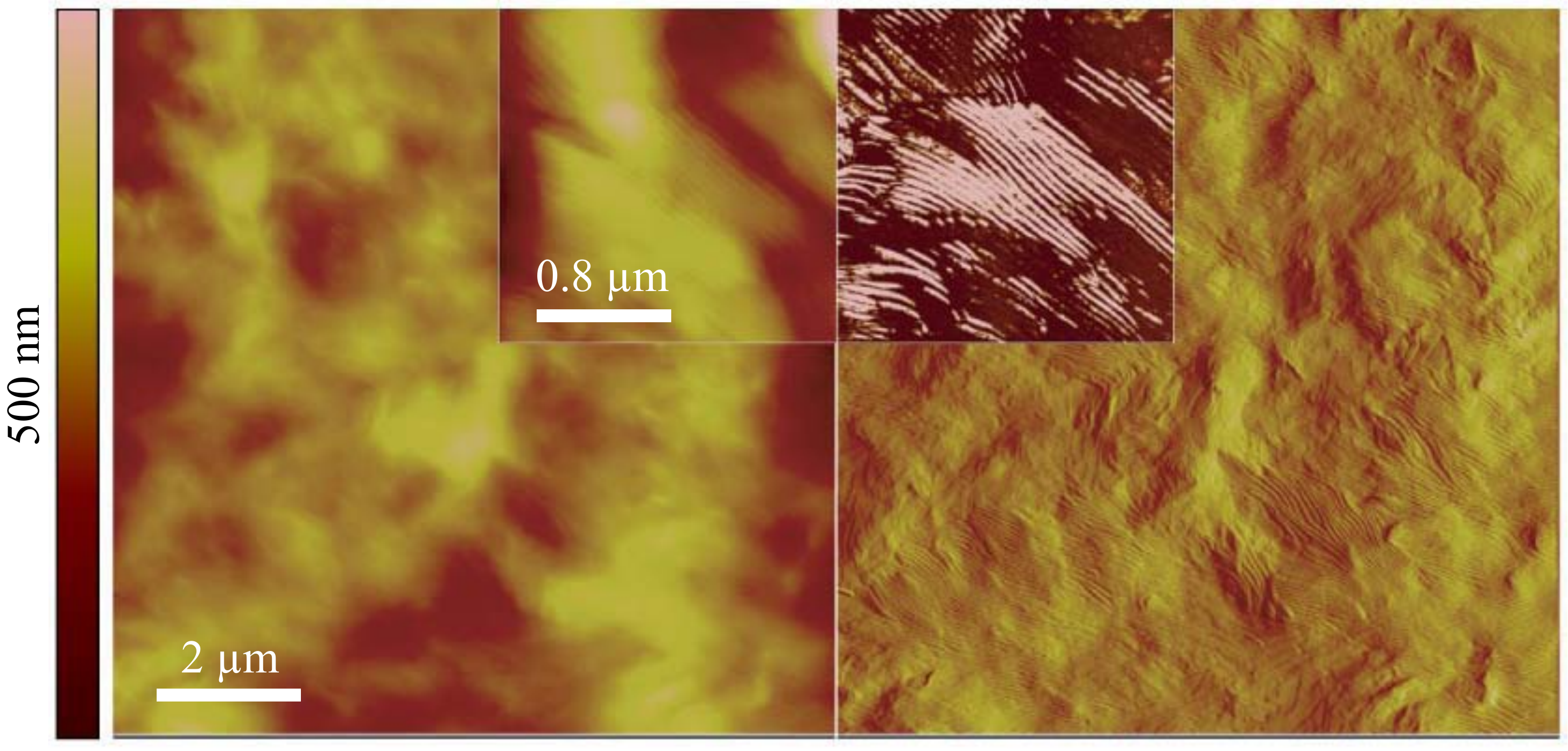}
\caption{AFM images of melt-cast polyethylene displaying ridges of crystallinity; the left side is a height-weighted image, and the right is the corresponding amplitude-weighted image better showing some fine details of surface morphology. The inset shows a height-weighted image on the left with a full height scale of 100~nm and the corresponding amplitude-weighted image on the right.} \label{fig_AFM1}	
\end{figure}

\begin{figure}
\centering
\includegraphics[width=6.0cm]{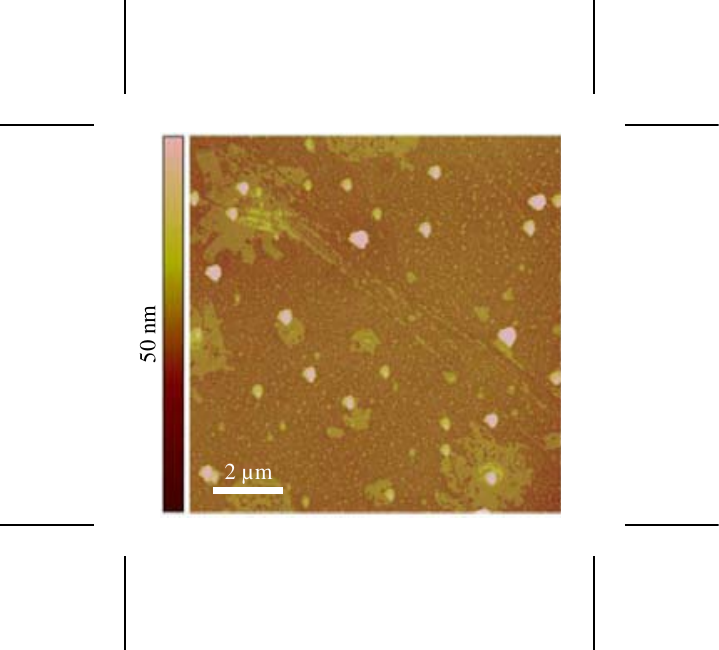}
\caption{Height-weighted AFM image of a broken cell containing pwMB after exposure to rubidium atoms, showing a lack of crystallinity and the possible presence of rubidium clusters.} \label{fig_AFM2}	
\end{figure}

Images taken of polyethylene are shown in Fig.~\ref{fig_AFM1}. We observe that surfaces of polyethylene and tetracontane melted onto the silicon substrate showed periodic ridges indicative of crystalline surface structure, consistent with the DSC results. In contrast, we show an image in Fig.~\ref{fig_AFM2} of a pwMB-coated glass surface after exposure to rubidium atoms; this surface was part of an operational magnetometer cell, coated in the standard manner described above, that was broken open for this experiment. This coating does not display any crystalline structure, again consistent with the DSC observations, and it features structures that may indicate either the presence of rubidium clusters on the surface or artifacts of such clusters having reacted upon exposure to air. Similar clusters have been observed on silane coatings \cite{Rampulla} and are speculated to represent regions of the film with an increased probability of causing alkali depolarization.

\subsection{Near edge X-ray absorption fine structure}
\label{sec:NEXAFS}

Near edge X-ray absorption fine structure (NEXAFS) spectroscopy was used to characterize the molecular bonds present in the paraffin. The NEXAFS experiments were carried out on the U7A NIST/Dow materials characterization end station at the National Synchrotron Light Source at Brookhaven National Laboratory. The X-ray beam was elliptically polarized (polarization factor of 0.85), with the electric field vector dominantly in the plane of the storage ring. The photon flux was 10$^{11}$~s$^{-1}$ at a typical storage ring current of 500~mA. A toroidal spherical grating monochromator was used to obtain monochromatic soft X-rays at an energy resolution of 0.2~eV. NEXAFS spectra were acquired for incident photon energy in the range 225-330~eV, which includes the carbon K edge. 
Each measurement was taken on a fresh spot of the sample in order to minimize possible beam damage effects.

\begin{figure*}
\centering
\includegraphics[width=13.0cm]{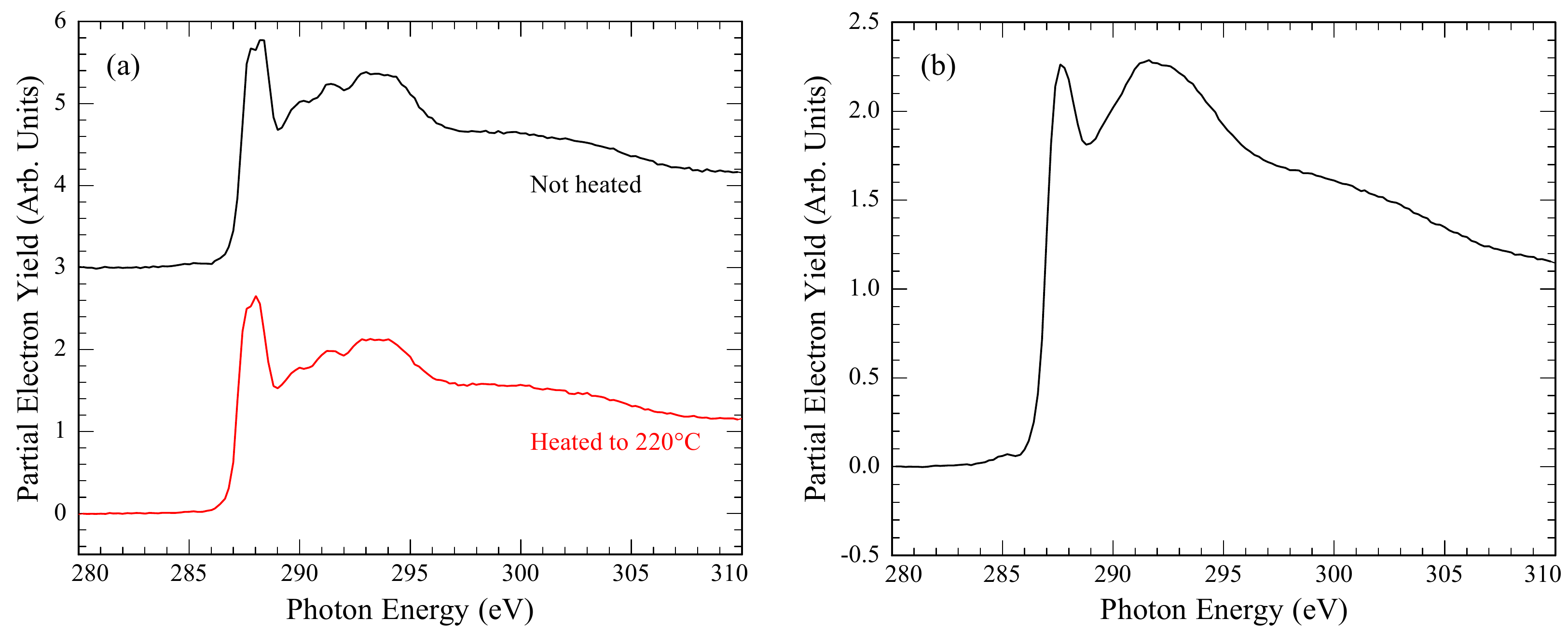}
\caption{(a)~NEXAFS spectra of tetracontane samples, displaying a C-H$^{*}$ peak near 288~eV and a $\sigma^{*}$~C--C peak near 292.3~eV. (b)~NEXAFS spectrum of polyethylene, displaying a small C=C peak near 285.3~eV.} \label{fig_NEXAFS1}	
\end{figure*}

The partial-electron-yield (PEY) signal was collected using a Channeltron electron multiplier with an adjustable entrance grid bias (EGB). All the data reported here are for a grid bias of -150~V. The channeltron PEY detector was positioned at an angle of 45$^{\circ}$ with respect to the incoming X-ray beam and off the equatorial plane of the sample chamber. To eliminate the effect of incident beam intensity fluctuations and monochromator absorption features, the PEY signals were normalized by the incident beam intensity obtained from the photo yield of a clean gold grid located along the path of the X-ray beam. A linear pre-edge baseline was subtracted from the normalized spectra, and the edge jump was arbitrarily set to unity at 320~eV, far above the carbon K-edge, a procedure that enabled comparison of different NEXAFS spectra for the same number of carbon atoms. Energy calibration was performed using a highly oriented pyrolytic graphite (HOPG) reference sample. The HOPG 1s to $\pi^{*}$ transition was assigned an energy of 285.3~eV according to the literature value. The simultaneous measurement of a carbon grid (with a 1s to $\pi^{*}$ transition of 285~eV) allowed the calibration of the photon energy with respect to the HOPG sample. Charge compensation was carried out by directing low-energy electrons from an electron gun onto the sample surface. Spectra are shown in Fig.~\ref{fig_NEXAFS1}(a) for tetracontane melted onto silicon, with one sample having been slowly heated to 220$^{\circ}$C over the course of about an hour prior to the measurement, featuring characteristic peaks due to a C-H$^{*}$ bond near 288~eV and a $\sigma^{*}$~C--C bond near 292.3~eV, in agreement with previous observations.\cite{HahnerNEXAFS}  The spectrum for polyethylene deposited on glass, shown in Fig.~\ref{fig_NEXAFS1}(b), displays these features as well as a small $\pi^{*}$ peak near 285.3~eV that is characteristic of a C=C double bond.\cite{StohrNEXAFS}

\begin{figure*}
\centering
\includegraphics[width=13.0cm]{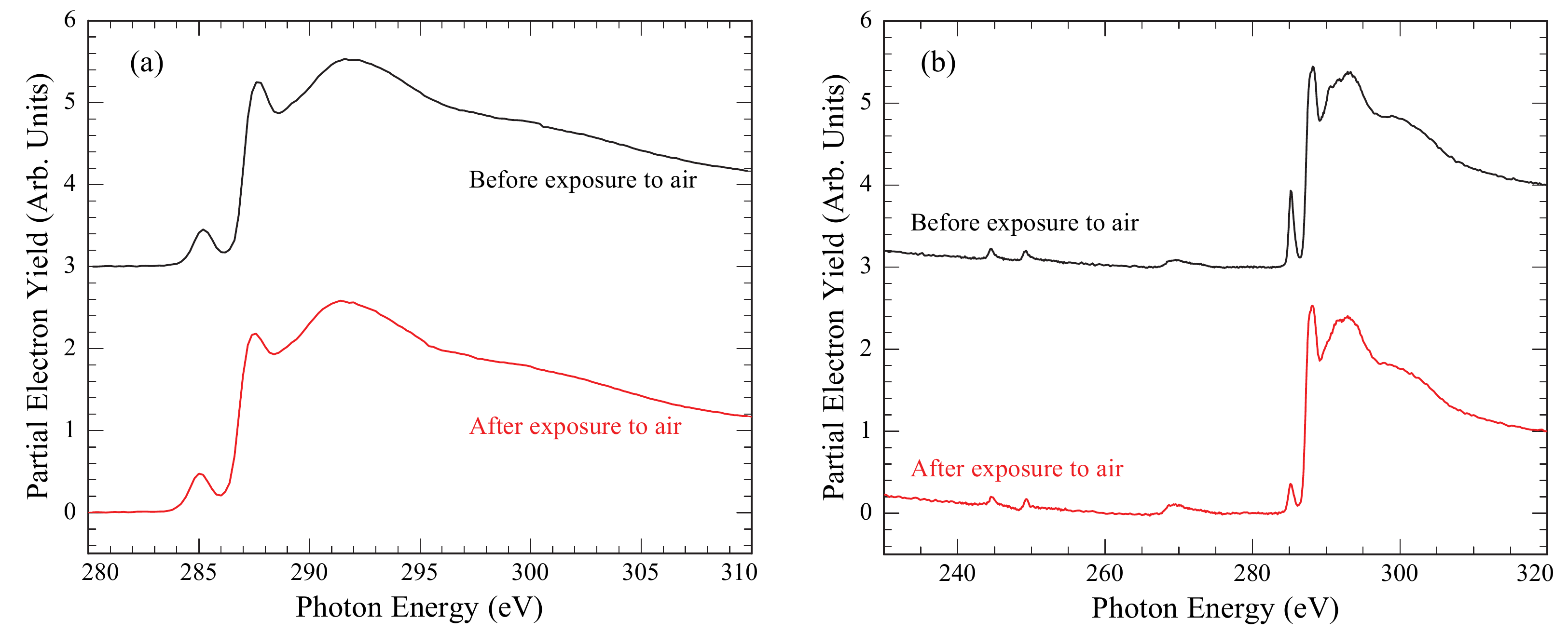}
\caption{NEXAFS spectra of pwMB samples both before and after exposure to air, showing a C=C double bond peak near 285.3~eV. (a)~Sample had not been exposed to alkali vapor. (b)~Sample had been exposed to cesium vapor, with the peaks at 244 and 249~eV assigned as the third harmonics of the cesium M5 and M4 edges, respectively.} \label{fig_NEXAFS2}	
\end{figure*}

Another experiment consisted of measuring the spectra from pwMB-coated glass slides, which were initially contained within glass cells; the standard coating procedure described above deposited pwMB on the slides in addition to the inside surface of the cells. In order to avoid exposure to air, the cells were broken open inside a glovebag containing an argon atmosphere, and the slides were then inserted into the load lock of the end station. After initial measurements were taken, the samples were exposed to air for several hours before conducting additional measurements. Figure~\ref{fig_NEXAFS2}(a) shows spectra of the surface of a pwMB sample that had not been exposed to alkali vapor, with the spectra appearing qualitatively similar both before and after exposure to air. In addition to the C-H$^{*}$ and C-C peaks, there is a peak at 285.3~eV due to a C=C double bond. These same features are evident in the spectra of a pwMB sample that had been exposed to cesium vapor for extended periods of time, shown in Fig.~\ref{fig_NEXAFS2}(b). The C=C peak is notably larger for the cesium-exposed sample before exposure to air, and is indeed larger than for the sample that had not been exposed to cesium; the reason for this is unknown but may be related to the ripening process, with the decrease in height after exposure to air likely due to oxidation, and additional measurements will be necessary to determine if the effect is repeatable. The peaks at approximately 244 and 249~eV are assigned as the third harmonics of the cesium M5 and M4 edges, respectively.\cite{CardonaLey} The C=C peak is also seen in the spectrum of the fractionated pwMB material melted directly onto glass (not shown).

\subsection{X-ray photoelectron spectroscopy}

Finally, X-ray photoelectron spectroscopy (XPS) was employed to determine the elemental composition of paraffin samples as well as to identify the chemical states of the elements present.   XPS is well-suited to a study of atomic vapor cells as it may be used to examine the nature of the alkali-paraffin interactions present in the cell.  In addition, angle-resolved XPS (ARXPS) offers a means to investigate the distribution of alkali metal in the coating as a function of depth.\cite{Rampulla}

XP spectra were acquired on a VG Scientific ESCALAB2 spectrometer with Al K$\alpha$ radiation ($h\nu$  = 1486.6~eV) and a base operating pressure of approximately 10$^{-9}$~Torr.  Survey scans were collected with a 1-eV step size, 100-ms dwell time, and 100-eV pass energy.  Higher-resolution scans collected with a 0.05-eV step size, 100-ms dwell time, and 20-eV pass energy were obtained for the C 1$s$, Rb 3$d$ and 3$p_{1/2}$/3$p_{3/2}$ regions.  Curve fitting of the core-level XPS lines was carried out with CasaXPS software using a nonlinear Shirley background subtraction and Gaussian-Lorentzian product function.  Tetracontane (Sigma-Aldrich; $>$98.5$\%$ purity) was coated onto piranha-cleaned (1:3 = 30$\%$ H$_{2}$O$_{2}$ :H$_{2}$SO$_{4}$; ca. 80$^{\circ}$C; 1~h) Si(100) substrates through immersion of 1-cm$^{2}$ wafers into melted paraffin wax at approximately 80$^{\circ}$C to give a visibly thick coating (ca. 100~$\mu$m).  Coated samples were placed in a cylindrical pyrex cell with a Rb source at one end (residual pressure $\sim$5x10$^{-7}$~Torr).  Exposure to Rb vapor was accomplished by heating the sealed pyrex cell to about 60$^{\circ}$C for 48~hours.  After trapping excess Rb vapor onto a room-temperature cold spot on the cell for 12~hours, the cell was broken open in air, and the samples were immediately transferred into the XPS antechamber.  Light exposure was minimized throughout the process in order to prevent light-induced desorption of the Rb atoms from the paraffin film.

\begin{figure}
\centering
\includegraphics[width=6.0cm]{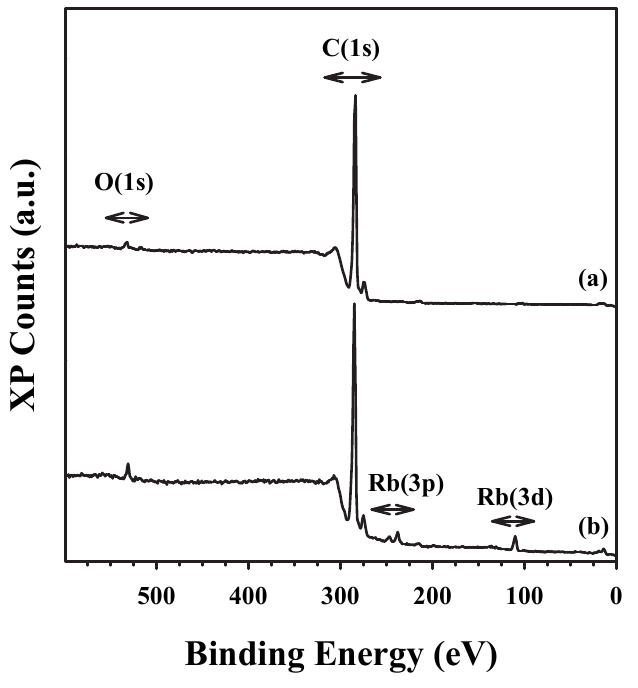}
\caption{XPS survey scans of (a) unexposed tetracontane and (b) tetracontane after exposure to Rb atoms, showing the appearance of Rb 3$d$ and 3$p_{1/2}$/3$p_{3/2}$ signals.}
\label{fig_XPS1}	
\end{figure}

\begin{figure*}
\centering
\includegraphics[width=14.0cm]{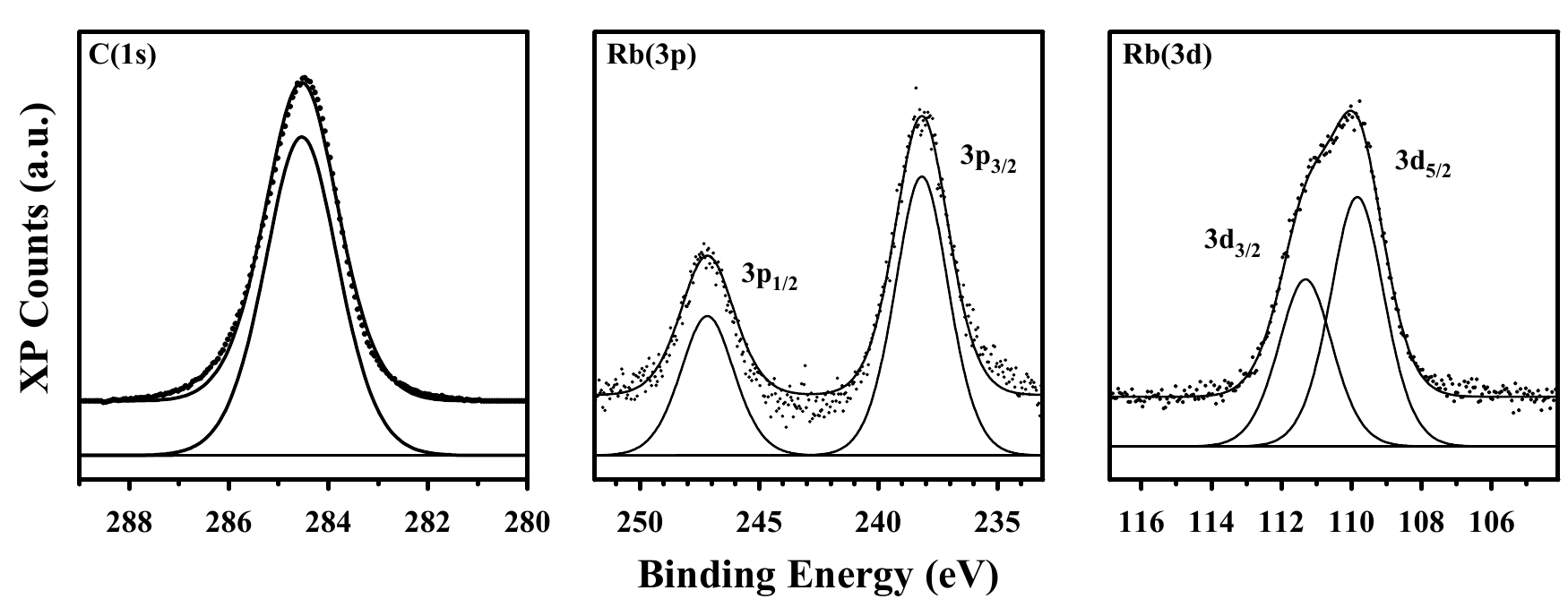}
\caption{XP spectra of C 1$s$, Rb 3$p$, and Rb 3$d$ core levels for Rb-exposed tetracontane.  Shown is the spectral envelope (solid line) and raw data above the fitted spectral peaks and background.  The symmetry of the C 1$s$ signal indicates the presence of a single carbon species on the surface.}
\label{fig_XPS2}	
\end{figure*}

XP spectra of both unexposed tetracontane and Rb-exposed tetracontane were collected.  Comparison of the survey scans (Fig.~\ref{fig_XPS1}) clearly shows the appearance of Rb 3$d$ (110~eV) and 3$p_{1/2}$/3$_{p3/2}$ (247/238~eV) signals in the Rb-exposed sample.  Notably, curve fitting of the C 1$s$ (284.5~eV) signal (Fig.~\ref{fig_XPS2}) indicates the presence of a single carbon species and suggests that there are no Rb-C bonds in this sample. Rb on the surface is merely physisorbed and not chemisorbed.

\begin{figure}
\centering
\includegraphics[width=6.0cm]{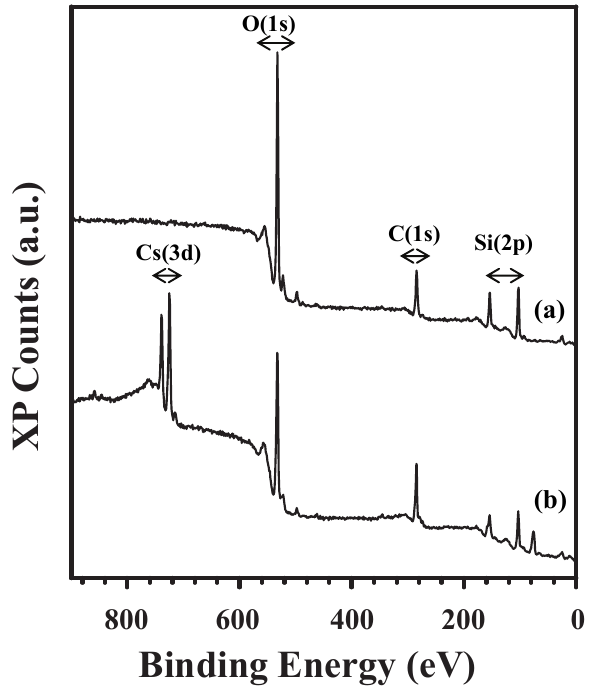}
\caption{XPS survey scans of (a) unexposed pwMB and (b) pwMB after exposure to Cs atoms, showing the appearance of Cs 3$d_{3/2}$/3$d_{5/2}$ signals.}
\label{fig_XPS3}	
\end{figure}

\begin{figure*}
\centering
\includegraphics[width=14.0cm]{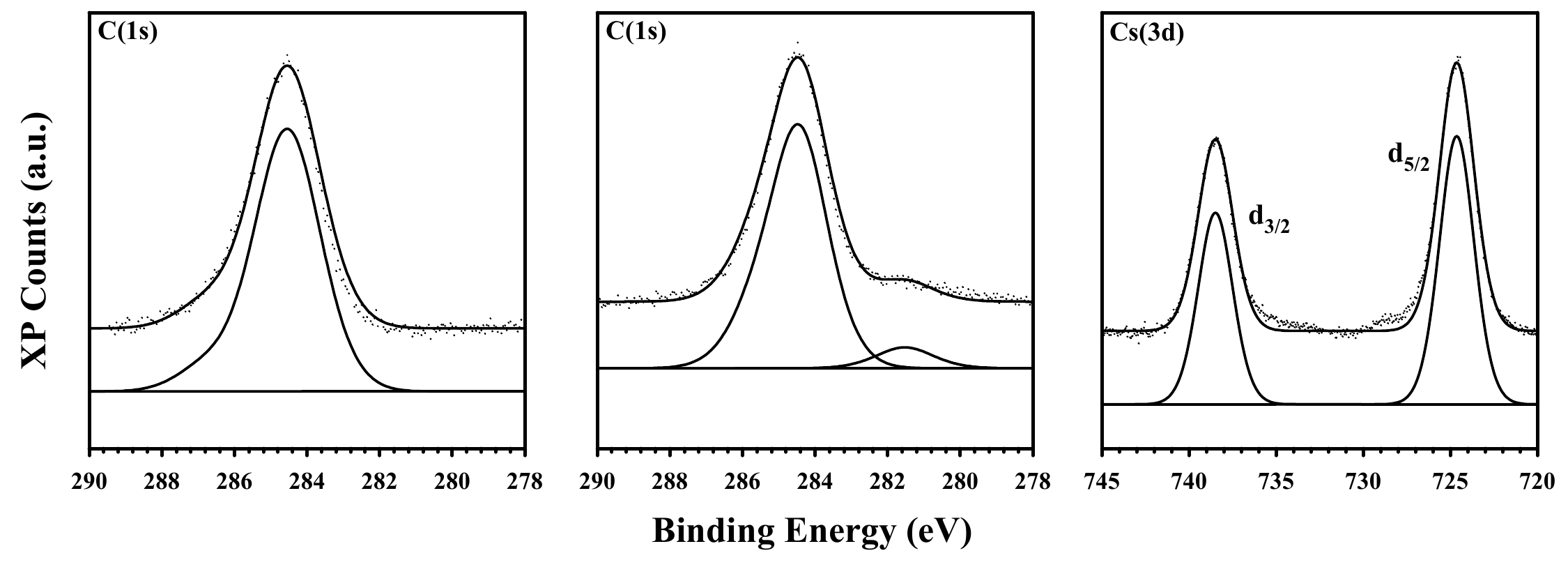}
\caption{From left to right: XP spectra of unexposed pwMB C 1$s$, and Cs-exposed pwMB C 1$s$ and Cs 3$d$ core levels.  Shown is the spectral envelope (solid line) and raw data above the fitted spectral peaks and background.  Notably, the Cs-exposed C 1$s$ region exhibits a second component assigned to a Cs-bound carbon species.}
\label{fig_XPS4}	
\end{figure*}

Analogous to the Rb-exposed tetracontane, samples of Cs-exposed pwMB clearly show the appearance of Cs signals in the XP spectrum, as shown in Fig.~\ref{fig_XPS3}. However, in contrast to the tetracontane samples, the Cs-exposed pwMB carbon peak exhibits an asymmetry that suggests the presence of more than one carbon species. Indeed, spectral deconvolution of the carbon region (Fig.~\ref{fig_XPS4}) indicates two components, with one at lower binding energy (ca. 281.5~eV) than the expected C-C/C-H signal at 284.5~eV. This signal is reproducible at different positions on the sample and does not appear in the unexposed pwMB sample, so it is probably not a result of surface charging effects. The lower binding energy component may be assigned to Cs-bound carbon \cite{MeyersXPS} and represents an alkali-carbon interaction not present in the Rb-exposed tetracontane. We speculate that this bound state arises from a reaction between Cs and C=C double bonds present in pwMB; such a pathway not only agrees with the observed C=C NEXAFS signal but also points to the importance of C=C double bonds in understanding the alkali-paraffin interactions.

\subsection{Light-Induced Atomic Desorption Observations}

In addition to the standard analytical techniques for studying surfaces and bulk materials, such as those described above, alkali vapor cell coatings have in the past typically been studied using more specialized physical methods, such as observations of Zeeman and hyperfine lifetimes and frequency shifts. Such studies permit direct comparison of the suitability of different coating materials for use in clocks and magnetometers. Along these lines, we describe here an experiment that allows comparison of the effect of desorbing light on alkali atoms absorbed into different types of paraffin coatings. The results of such an experiment, combined with the information provided by the other tests described above, can guide the design of coating materials for LIAD-loaded atomic devices.

For this experiment, we use two different cell geometries. The first geometry features lockable stems to prevent ejected atoms from leaving the main cell body. The lock is implemented using a sliding glass ``bullet'' and permits large vapor density changes to be maintained after repeated exposures to desorbing light; for more details on the experimental setup and photographs of cells with this lockable stem design, see Ref.~\onlinecite{KaraulanovThermal}. All such cells are spherical with diameter of $3$~cm, and the stems remain locked at all times. The second geometry features cylindrical cells, approximately 9~cm long and 2~cm in diameter, without stem locks. All cells of either geometry contain rubidium in natural abundance. For the first geometry, the atomic density is determined by monitoring the absorption of a weak (1~$\mu$W) probe laser beam tuned to the $\textrm{F}_{g}=3 \rightarrow \textrm{F}_{e}$ transitions of the $^{85}$Rb $\textrm{D}_{2}$ line. For the second geometry, the density is determined by quickly sweeping the probe beam across all $\textrm{F}_{g} \rightarrow \textrm{F}_{e}$ transitions of the $^{85}$Rb and $^{87}$Rb $\textrm{D}_{2}$ lines and fitting the measured transmission spectrum at each point in time. In order to induce atomic desorption, the cells are fully illuminated by off-resonant light produced by a 405~nm (blue) laser diode. Cells are characterized by the LIAD yield $\eta=(n_{\textnormal{max}}-n_{0})/n_{0}$, where $n_{\textnormal{max}}$ is the maximum Rb vapor density measured after exposure to desorbing light, and $n_{0}$ is the initial density prior to illumination.

\begin{table*}
\begin{center}
\begin{tabular}{|l|c|c|c|} \hline
Coating Material&Preparation Temperature ($^{\circ}$C)&LIAD Yield (1)&LIAD Yield (2)\\
\hline\hline
Alkene 80&175& &$<0.01$\\
\hline
pwMB&260& $0.6$&$0.13$\\
\hline
FR-130&260&$<0.1$&$<0.01$\\
\hline
Alkene 110&270& & $0.07$\\
\hline
ENET4160&275& &$<0.01$\\
\hline
$C_{40}H_{82}$ (Tetracontane)&280&$<0.05$&$0.17$\\
\hline
$C_{44}H_{90}$ (Tetratetracontane)&290&$<0.08$&$<0.01$\\
\hline
Deuterated polyethylene&320&$0.7$&$0.29$\\
\hline
Deuterated polyethylene&360&$8$&$1.6$\\
\hline
\end{tabular}
\end{center}
\caption{LIAD yield $\eta$ for several different paraffin and alkene materials in cells of the two geometries described in the text. The 405~nm desorbing light had intensity of 5~mW/cm$^{2}$ for cells of the first geometry and 2~mW/cm$^{2}$ for cells of the second geometry. The listed temperature is that at which the coating was deposited on the inside surface of the cell.}
\label{table_LIAD}
\end{table*}

Results of the LIAD experiment are summarized in Table~\ref{table_LIAD}, with an error in the determination of $\eta$ of approximately 15\% for each cell. The preparation temperature is given at which each material was coated on the inside surface of its cell. For the data shown, desorbing light intensity of 5~mW/cm$^{2}$ was used with the cells of the first geometry, and intensity of 2~mW/cm$^{2}$ was used with cells of the second geometry. In addition to paraffin materials, several alkene materials are also considered. The materials labeled ``Alkene~80'' and ``Alkene~110'' are produced by fractionation of Alpha Olefin Fraction C20-24 from Chevron Philips Chemical Company, which contains a mixture of straight-chain alkenes, with primarily between 18 and 26 carbon atoms; the number in the material label refers to the temperature of distillation. Alkene~80 is the highly effective anti-relaxation coating material described in Ref.~\onlinecite{BalabasMagic}. ENET4160 from Gelest, Inc. contains 1-triacontene and small amounts of other alkenes with between 26 and 54 carbon atoms.

\begin{figure}
\centering
\includegraphics[width=8.0cm]{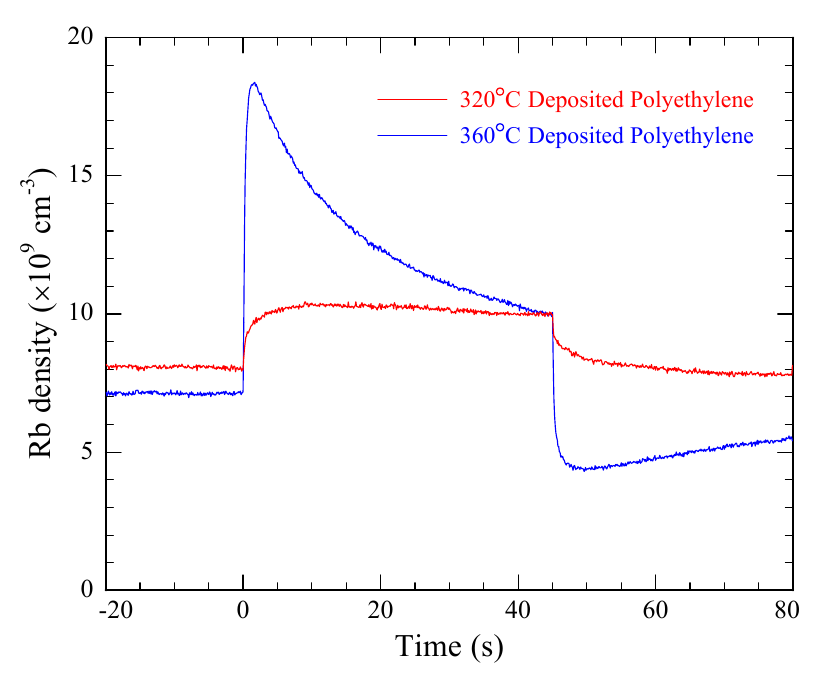}
\caption{Measured change in Rb vapor density due to application of desorbing light in two cells of the second geometry, both coated with deuterated polyethylene, but with the coating deposited at different temperatures. The desorbing light is turned on at $t$=0~s and turned off at $t$=45~s.} \label{fig_LIAD}	
\end{figure}

The LIAD yield is highly dependent on cell geometry and can not be directly compared between samples of different geometries, even when accounting for differing intensity of the desorbing light. Samples of pwMB and deuterated polyethylene show appreciable increase in rubidium density after exposure to desorbing light for both cell geometries, as does Alkene~110 for the second geometry. It is interesting to note that the NEXAFS spectra of both pwMB and polyethylene show evidence of unsaturated C=C bonds. Although Alkene~80 and ENET4160 do not display any measurable LIAD effect, these results nevertheless suggest that covalently bound alkali atoms could act in part as a reservoir for the LIAD effect. Tetracontane also shows a large LIAD yield, but only for the second experiment. For both cell geometries, deuterated polyethylene gives a larger LIAD yield when the material is deposited at higher temperature, and it also presents very different dynamical behavior depending on the preparation temperature, as shown in Fig.~\ref{fig_LIAD}; after the desorbing light turns off, the vapor density for the sample prepared at 360$^{\circ}$C is observed to be less than the initial density due to depletion of atoms from the coating.\cite{AleksandrovLIAD} These results are preliminary and motivate a more comprehensive study of cell-to-cell variation and the effects of preparation conditions (including particularly temperature), material deuteration, unsaturated bonds, and other coating parameters.

\section{Discussion and Conclusions}

These measurements provide important details about the properties of effective anti-relaxation coatings and demonstrate the utility of applying surface-science methods. For instance, paraffin does not need to be in crystalline form, as evidenced by the observations of the pwMB wax using DSC and AFM. Particularly interesting is the detection of C=C double bonds in the pwMB wax using NEXAFS, a surface technique, while transmission FTIR, a bulk measurement technique, does not show any C=C bonds above the detection limit of approximately 5\%; the discrepancy may be due to lack of sensitivity in the FTIR measurement, but it is also possible that the molecules with unsaturated bonds represent impurities expelled to the surface of the material and not extant in its bulk. It is unknown if the double bonds are present in the raw material or if they form during fractionation when heated to 220$^{\circ}$C. Alkanes are known to decompose at high temperatures,\cite{Madorsky, JawDecomposition} potentially leading to the presence of unsaturated alkenes in the deposited pwMB surface coating, although the NEXAFS spectrum of tetracontane shown in Fig.~\ref{fig_NEXAFS1} does not display evidence of double-bond formation after heating to 220$^{\circ}$C. Decomposition could explain the observed increase in LIAD yield for polyethylene materials deposited at higher temperatures. The alkene coating described in Ref.~\onlinecite{BalabasMagic} also contains double bonds and allows approximately 10$^{6}$ bounces with the surface, significantly more than any other known material. The presence of C=C double bonds in effective anti-relaxation coatings is unexpected because unsaturated bonds increase the polarizability of the surface, and effective coatings have long been assumed to require low polarizability to enable short alkali atom residence time, although these recent results indicate that this assumption may be mistaken. The unsaturated double-bond sites may also react with alkali atoms to form alkali-carbon bonds within the coating material, as detected in the pwMB samples using XPS. In fact, it is likely that passivation of the double-bond sites near the surface of the material is a necessary part of the ripening process.

We may also compare the DSC observations of the thermal properties of the materials with prior measurements of the temperature dependence of the wall shift in paraffin-coated cells of the rubidium 0-0 hyperfine frequency, which is used for frequency reference in atomic clocks. The hyperfine frequency in a cell coated with tetracontane exhibits hysteresis:\cite{RahmanJQE} the frequency changes little until the solid tetracontane is heated to about 80$^{\circ}$C, at which point it decreases significantly as the wax melts, and the frequency does not increase again upon cooling until the liquid tetracontane reaches a temperature 1$^{\circ}$C colder than this and resolidifies. This behavior agrees well with the measured temperatures of the melting and fusion peaks of tetracontane. In addition, the temperature dependence of the hyperfine frequency in cells coated with paraflint is observed to change sign at 72$^{\circ}$C,\cite{Brewer} which correlates with the endothermic peak measured with DSC.

Our observations suggest several avenues for further research. Specific surface characteristics such as roughness and fractional coverage can be studied in order to optimize the composition of coating materials and the deposition procedure. The absence or presence of alkali-carbon bonds has significant implications with regard to the mechanism of the ripening and LIAD processes, and so future work will focus on the effect of unsaturated bonds on both the anti-relaxation effectiveness and LIAD efficiency of materials. For example, angle-resolved XPS studies can reveal depth-dependent changes in the distribution of alkali and alkali-bound species in paraffins before and after treatment with desorbing light. This work can also be extended to include additional techniques for the study of coated surfaces, such as sum frequency generation (SFG) spectroscopy \cite{ChenShen} and Raman spectroscopy;\cite{Schrader} the latter may be particularly useful for its ability to observe the surfaces of intact, operational vapor cells.

In conclusion, using modern analytic techniques we show a systematic study of paraffin waxes which can be extended to a broader set of waxes and other anti-relaxation surface coatings to understand the fundamental properties of these coatings. Combining a number of different surface science methods gives information regarding the chemical nature of the bulk coating materials as well as the thin films at the surface. This knowledge will inform the design of coatings used with atomic devices. As this research continues, it will hopefully lead to the development of more effective and robust anti-relaxation coatings for use in a variety of alkali-vapor-based technologies under a wide range of operational conditions.

\section*{Acknowledgments}

The authors thank Daniel Fischer, Kristin Schmidt, and Ed Kramer for assistance with the NEXAFS measurements, and Joel Ager, Joshua Wnuk, David Trease, and Gwendal Kervern for helpful discussions and other assistance. SJS, DJM, MHD, AP, and DB, the Advanced Light Source, and the DSC, FTIR, and AFM studies were supported by the Director, Office of Science, Office of Basic Energy Sciences, Materials Sciences Division and Nuclear Science Division, of the U.S. Department of Energy under Contract No. DE-AC02-05CH11231 at Lawrence Berkeley National Laboratory. Other parts of this work were funded by NSF/DST Grant No. PHY-0425916 for U.S.-India cooperative research, by an Office of Naval Research (ONR) MURI grant, and by ONR Grant N0001409WX21049.

Certain commercial equipment, instruments, or materials are identified in this document. Such identification does not imply recommendation or endorsement by the U.S. Department of Energy, Lawrence Berkeley National Laboratory, the Advanced Light Source, or the National Institute of Standards and Technology, nor does it imply that the products identified are necessarily the best available for the purpose.

\bibliographystyle{prsty}

\end{document}